\documentclass[12pt]{article}

\usepackage{axodraw}
\usepackage{epsf}
\usepackage{rotating}

\catcode`@=11
\def\citer{\@ifnextchar
[{\@tempswatrue\@citexr}{\@tempswafalse\@citexr[]}}

\def\@citexr[#1]#2{\if@filesw\immediate\write\@auxout{\string\citation{#2}}\fi
  \def\@citea{}\@cite{\@for\@citeb:=#2\do
    {\@citea\def\@citea{--\penalty\@m}\@ifundefined
       {b@\@citeb}{{\bf ?}\@warning
       {Citation `\@citeb' on page \thepage \space undefined}}%
\hbox{\csname b@\@citeb\endcsname}}}{#1}}
\catcode`@=12

\newcommand{\beq}{\begin{eqnarray}}
\newcommand{\eeq}{\end{eqnarray}}
\newcommand{\nn}{\noindent}
\newcommand{\non}{\nonumber}

\newcommand{\ra}{\rightarrow}
\newcommand{\s}{\\ \vspace*{-3mm} } 
\newcommand{\tgb}{\tan\beta}

\newcommand{\lsim}{\raisebox{-0.13cm}{~\shortstack{$<$ \\[-0.07cm] $\sim$}}~}

\begin{document}

\def\thefootnote{\fnsymbol{footnote}}

\begin{flushright}
DESY 99--196 \\
PM/99--60\\
hep-ph/9912476 \\
December 1999
\end{flushright}

\vspace{1cm}

\begin{center}

{\large\sc {\bf SUSY--QCD Corrections to Higgs Boson}}  

\vspace*{3mm}

{\large\sc {\bf  Production at Hadron Colliders}} 

\vspace{1cm}

{\sc A. Djouadi$^1$ and M. Spira$^2$\footnote{Heisenberg-Fellow}}

\vspace{0.5cm}

$^1$ Laboratoire de Physique Math\'ematique et Th\'eorique, UMR5825--CNRS,\\
Universit\'e de Montpellier II, F--34095 Montpellier Cedex 5, France.

\vspace*{0.3cm}

$^2$ II. Institut f\"ur Theoretische Physik\footnote{Supported by
the EU FF Programme under contract FMRX-CT98-0194 (DG 12 - MIHT)},
Universit\"at Hamburg, \\
Luruper Chaussee 149, D--22761 Hamburg, Germany. 

\end{center}

\vspace{2cm}

\begin{abstract}
\nn
We analyze the next-to-leading order SUSY-QCD corrections to the
production of
Higgs particles at hadron colliders in supersymmetric extensions of the
Standard Model. Besides the standard QCD corrections due to gluon exchange and
emission, genuine supersymmetric corrections due to the virtual exchange of
squarks and gluinos are present. At both the Tevatron and the LHC, these 
corrections are found to be small in the Higgs-strahlung, Drell--Yan-like
Higgs pair production and vector boson fusion processes. 
\end{abstract}

\newpage

\def\thefootnote{\arabic{footnote}}
\setcounter{footnote}{0}

\subsection*{1. Introduction}

A firm prediction of supersymmetric extensions of the Standard Model (SM)
\cite{R0} is the existence of a light scalar Higgs boson. In the Minimal
Supersymmetric Standard Model (MSSM) the Higgs sector contains a quintet of
scalar particles [two CP-even $h$ and $H$, a pseudoscalar $A$ and two charged
$H^\pm$ particles] \cite{R1}, the Higgs boson $h$ of which should be light,
with a mass $M_{h} \lsim 135$ GeV \cite{R2}. If this particle is not found at
LEP2 \cite{lep}, it will be produced at the upgraded Tevatron (where a large
luminosity, $\int {\cal L} \sim 20$ fb$^{-1}$, is expected) \cite{tev} or at
the LHC \cite{lhc,habil}, if the MSSM is indeed realized in Nature. \s

At hadron colliders, the dominant production mechanism for neutral Higgs
particles is the (heavy quark) loop induced gluon fusion process, $gg 
\ra \Phi$ with $\Phi=h,H$ or $A$ \cite{ggh}. Since the Higgs
particles in the mass range of interest, $M_\Phi \lsim 135$ GeV, dominantly
decay into bottom quark pairs, this process is rather difficult to exploit
at the Tevatron
because of the huge QCD background \cite{tev}.  In contrast, at the LHC rare
decays of the lightest $h$ boson to two photons or decays of the heavy $H,A$
boson to $\tau$ and $\mu$ lepton pairs make this
process very useful \cite{lhc}.  \s

Additional Higgs production mechanisms at hadron colliders are provided by: \s

{\bf a)} Higgs-strahlung off $W$ or $Z$ bosons for the CP-even Higgs particles
[due to CP-invariance, the pseudoscalar $A$ particle does not couple to the
massive gauge bosons at tree level]: $q\bar{q} \ra V^* \ra \Phi V$ with
$\Phi=h,H$ and $V=W,Z$ \cite{vh}. At the Tevatron, the process $q\bar{q}' \ra
hW$ [with the $h$ boson decaying into $b\bar{b}$ pairs] develops a cross
section of the order of a fraction of a picobarn for a SM-like $h$ boson with a
mass below $\sim 135$ GeV, making it the most relevant mechanism to study
\cite{tev}. At the LHC, both the $b\bar{b}$ and $\gamma \gamma$ decay modes of
the $h$ boson may be exploited \cite{lhc}. \s

{\bf b)} If the heavier $H,A, H^\pm$ bosons are not too massive, the pair 
production of two Higgs particles in the Drell--Yan type process, $q\bar{q}
\ra \Phi_1 \Phi_2$ \citer{pair,krause}, might lead to a variety of final 
states [$hA, HA, H^\pm h, H^\pm H, H^\pm A, H^+ H^-$] with reasonable cross 
sections [in particular for $M_A \sim M_H \sim M_{H^\pm} \lsim 250$ GeV and 
small values of tan$\beta$, the ratio of the vacuum expectation values of 
the two Higgs doublets] especially at the LHC. Moreover, neutral and
charged Higgs boson pairs will be produced in gluon fusion $gg\to \Phi_1\Phi_2$
\citer{9a,9b}. \s

{\bf c)} The production of CP-even Higgs bosons via vector boson fusion,
$q q \ra qqV^*V^* \ra qq\Phi$ \cite{vvh}: In the case of a SM-like $h$ boson,
this process has a sizeable cross section at the LHC.
While decays of the Higgs boson into heavy quark pairs are problematic to be
detected in the jetty environment of the LHC, decays into $\tau$ lepton
pairs make this process useful at the LHC \cite{zepp}. \s

In addition to these types of processes, neutral Higgs boson radiation off 
heavy bottom and top quarks [$q\bar{q},gg\ra b\bar{b}\Phi,
t\bar{t}\Phi$] might play an important role in SUSY theories \cite{tth}.
In particular, because the couplings of the Higgs boson to $b$ quarks can be
strongly enhanced for large values of tan$\beta$, Higgs boson production in 
association with $b\bar{b}$ pairs can give rise to large production
rates. \s

It is well known that for processes involving strongly interacting particles,
as is the case for the ones discussed above, the lowest order cross 
sections are affected by large uncertainties arising from higher
order corrections. If the next-to-leading QCD corrections to these
processes are included, the total cross sections can be defined properly 
and in a reliable way in most of the cases. \s

For the standard QCD corrections, the next-to-leading corrections are available
for most of the Higgs boson production processes. The K-factors [defined as the
ratios of the next-to-leading order cross sections to the lowest order ones]
for Higgs boson production via the gluon fusion processes have been calculated
a few years ago \cite{gghqcd}; the [two-loop] QCD corrections have been found
to be significant since they increase the cross sections by up to a factor of
two. The K-factors for Higgs production in association with a gauge boson $(a)$
and for Drell--Yan-like Higgs pair production $(b)$, can be inferred from the
Drell--Yan production of weak vector bosons and increase the cross section by
approximately 30\% \cite{vhqcd}.  The QCD corrections to $gg\to \Phi_1\Phi_2$
are only known in the limit of light Higgs bosons compared with the loop-quark
mass; they enhance the cross sections by up to a factor of two \cite{9b}.  For
Higgs boson production in the weak boson fusion process $(c)$, the QCD
corrections can be derived in the structure function approach from
deep-inelastic scattering; they turn out to be rather small, enhancing the
cross section by about 10\% \cite{vvhqcd}. Finally, the full QCD corrections to
the associated Higgs production with heavy quarks are not yet available; they
are only known in the limit of light Higgs particles compared with the heavy
quark mass \cite{tthqcd} which is only applicable to $t\bar{t}h$ production; in
this limit the QCD corrections increase the cross section by about 20--60\%. \s

Besides these standard QCD corrections, additional SUSY-QCD corrections must be
taken into account in SUSY theories; the SUSY partners of quarks and gluons,
the squarks and gluinos, can be exchanged in the loops and contribute to the
next-to-leading order total cross sections. In the case of the gluon fusion
process, the QCD corrections to the squark loop contributions have been
calculated in the limit of light Higgs bosons and heavy gluinos; the K-factors
were found to be of about the same size as the ones for the quark loops
\cite{gghqcdsusy}. The SUSY--QCD corrections to the other production processes
are not yet available. \s

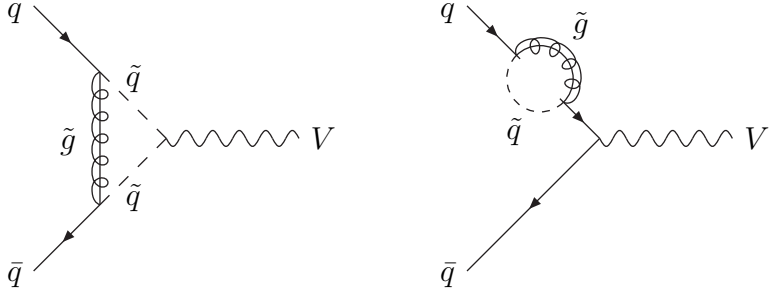
\begin{figure}[hbt]
\begin{center}
\begin{picture}(180,100)(-50,0)

\ArrowLine(0,100)(25,75)
\ArrowLine(25,25)(0,0)
\Line(25,75)(25,25)
\Gluon(25,75)(25,25){-3}{5}
\DashLine(25,75)(50,50){5}
\DashLine(50,50)(25,25){5}
\Photon(50,50)(100,50){-3}{5}
\put(105,46){$V$}
\put(-10,96){$q$}
\put(-10,-4){$\bar q$}
\put(35,70){$\tilde{q}$}
\put(35,25){$\tilde{q}$}
\put(10,46){$\tilde{g}$}

\end{picture}
\begin{picture}(180,100)(-30,0)

\ArrowLine(0,100)(20,80)
\ArrowLine(35,65)(50,50)
\ArrowLine(50,50)(0,0)
\Photon(50,50)(100,50){-3}{5}
\GlueArc(27.5,72.5)(12.5,-45,135){3}{4}
\CArc(27.5,72.5)(12.5,-45,135)
\DashCArc(27.5,72.5)(12.5,135,315){3}
\put(105,46){$V$}
\put(-10,96){$q$}
\put(-10,-4){$\bar q$}
\put(15,50){$\tilde{q}$}
\put(40,90){$\tilde{g}$}

\end{picture}
\caption[ ]{\label{fg:dia} \it Generic diagrams contributing to the SUSY-QCD corrections to
the $q\bar q V$ vertex [$V=\gamma,Z,W$] at NLO.}
\end{center}
\end{figure}

In this paper, we calculate the SUSY--QCD corrections to the Higgs production
cross sections for Higgs-strahlung, Drell--Yan-like Higgs pair production
and weak boson 
fusion processes. These corrections originate from $q\bar{q}V$ one-loop 
vertex corrections, where squarks of the first two generations and gluinos
are exchanged,
and the corresponding quark self-energy counterterms, Fig.~\ref{fg:dia}.
These corrections
turn out to be small for reasonably large values of the squark and
gluino masses, altering the cross section by a few percent only. For 
heavier masses, squarks and gluinos decouple and the SUSY-QCD corrections 
become tiny.
 \s

Note that genuine SUSY--QCD corrections are also present in the decays of the
Higgs bosons into heavy quark pairs \cite{decays}. They are much larger
than for the production in the case of  large mixing in the sbottom sector,
being of the order of a few ten per cent, and decouple only slowly 
for large squark and gluino masses. They are being included in the program 
HDECAY \cite{hdecay} which calculates the decay widths and branching ratios of 
the MSSM Higgs particles.  

\subsection*{2. QCD corrections to the production processes}

\subsubsection*{2.1 Higgs-strahlung and Drell--Yan-like pair production
processes}

At hadron colliders, the lowest order partonic cross section for the 
Higgs-strahlung processes, $q\bar{q} \ra V\Phi$ with $V=W,Z$ and
$\Phi=h,H$, is given by \cite{habil,vh} 
\beq
\hat{\sigma}_{\rm LO}(q\bar{q} \ra V \Phi)= \frac{G_F^2 M_V^4}{288 \pi \hat{s}}
g_{\Phi VV}^2 (v_q^2 + a_q^2) \lambda^{1/2} (M_V^2, M_\Phi^2; \hat{s}) \frac{
 \lambda(M_V^2, M_\Phi^2; \hat{s})+12 M_V^2/\hat{s}}{(1-M_V^2/\hat{s})^2}
\eeq
with the couplings $g_{\Phi VV}=\sin(\beta-\alpha)$ or $\cos(\beta-\alpha)$ 
for $h$ and $H$ respectively; $\hat{s}$ is the partonic c.m. energy and 
$\lambda$ the usual two-body phase space function $\lambda(x,y;z)=(1-x/z-y/z)
^2-4xy/z^2$. $v_q, a_q$ are the vector and axial-vector couplings of the quark
$q$ to vector bosons and are given by $v_q=2I_{q}^3-4e_q s_W^2, 
a_q=2I_q^3$ for $V=Z$ [$e_q$ is the electric charge, $I_q^3$ the weak isospin
of  the quark and $s_W^2 =1-c_W^2 \equiv \sin^2 \theta_W$] and $v_q=a_q=
\sqrt{2}$ for $V=W$. \s

The partonic cross section for Drell--Yan-like Higgs pair production $q\bar{q}
\ra \Phi_1 
\Phi_2$, where at least one of the Higgs bosons is neutral, is given by
\cite{pair,9a}: 
\beq
\hat{\sigma}_{\rm LO}(q\bar{q} \ra \Phi_1 \Phi_2)= \frac{G_F^2 M_V^4}{288 \pi 
\hat{s}} g_{\Phi_1 \Phi_2 V}^2 (v_q^2 + a_q^2) \frac{ \lambda^{3/2} (
M_{\Phi_1}^2 ,M_{\Phi_2}^2; \hat{s})} {(1-M_V^2/\hat{s})^2}
\eeq
where the couplings are given by: $g_{ZhA}= g_{W h H^+} =\cos(\beta-\alpha)$, 
$g_{ZHA}= g_{W H H^+} =\sin(\beta-\alpha)$ and $g_{WH^+A}=1$. For charged
Higgs boson pair production there is an $s$-channel photon exchange in addition
to the $Z$-channel diagram, and the cross section is given by
\cite{pair,krause}:
\beq
\hat{\sigma}_{\rm LO}(q\bar{q} \ra H^+H^-) = \frac{\pi \alpha^2(\hat{s})}{9
\hat{s} }  \bigg[ e_q^2 + \frac{2 e_q v_q v_{H^+}}{16 s_W^2 c_W^2 (1- M_Z^2/ 
\hat{s})} + \frac{(v_q^2+a_q^2)v_{H^+}^2}{ 64 s_W^4 c_W^4 (1-
M_Z^2/\hat{s})^2}  \bigg]
\eeq
with $v_{H^+} =-2 +4s_W^2$.

The hadronic cross sections can be obtained from convoluting eqs.~(1--3) 
with the corresponding (anti)quark densities of the protons 
\begin{eqnarray}
\sigma_{LO} (pp \ra V\Phi, \Phi_1 \Phi_2) = \int_{\tau_0}^1 d\tau 
\sum_q \frac{d{\cal L}^{q\bar q}}{d\tau} \hat\sigma_{LO}(\hat{s}=\tau s)  
\end{eqnarray}
where $\tau_0=(M_V+M_\Phi)^2/s$ for the Higgs-strahlung and $\tau_0= 
(M_{\Phi_1}+M_{\Phi_2})^2/s$ for the pair production processes, with $s$ being 
the total hadronic c.m.~energy squared. \s

The standard QCD corrections, with virtual gluon exchange, gluon emission 
and quark emission,
are identical to the corresponding corrections to 
the Drell--Yan process \cite{vhqcd,DY}. They modify the lowest order cross 
section in the following way \cite{habil,vhqcd}
\begin{eqnarray}
\sigma & = & \sigma_{LO} + \Delta\sigma_{q\bar q} + \Delta\sigma_{qg}
\nonumber \\
\Delta\sigma_{q\bar q} & = & \frac{\alpha_s(\mu)}{\pi} \int_{\tau_0}^1 d\tau
\sum_q \frac{d{\cal L}^{q\bar q}}{d\tau} \int_{\tau_0/\tau}^1 dz~\hat
\sigma_{LO}(Q^2 = \tau z s)~\omega_{q\bar q}(z) \nonumber \\
\Delta\sigma_{qg} & = & \frac{\alpha_s(\mu)}{\pi} \int_{\tau_0}^1 d\tau
\sum_{q,\bar q} \frac{d{\cal L}^{qg}}{d\tau} \int_{\tau_0/\tau}^1 dz~\hat
\sigma_{LO}(Q^2 = \tau z s)~\omega_{qg}(z)
\end{eqnarray}
with the coefficient functions \cite{DY}
\begin{eqnarray}
\omega_{q\bar q}(z) & = & -P_{qq}(z) \log \frac{M^2}{\tau s}
+ \frac{4}{3}\left\{ \left[\frac{\pi^2}{3} -4\right]\delta(1-z) +
2(1+z^2)\left(\frac{\log(1-z)}{1-z}\right)_+ 
\right\} \nonumber \\
\omega_{qg}(z) & = & -\frac{1}{2} P_{qg}(z) \log \left(
\frac{M^2}{(1-z)^2 \tau s} \right) + \frac{1}{8}\left\{ 1+6z-7z^2 \right\} \, ,
\end{eqnarray}
where $M$ denotes the factorization scale, $\mu$ the renormalization scale
and $P_{qq}, P_{qg}$  the well-known Altarelli--Parisi splitting functions, 
which are given by \cite{apsplit}
\begin{eqnarray}
P_{qq}(z) & = & \frac{4}{3} \left\{ \frac{1+z^2}{(1-z)_+}+\frac{3}{2}\delta(1-z)
\right\} \nonumber \\
P_{qg}(z) & = & \frac{1}{2} \left\{ z^2 + (1-z)^2 \right\} \, .
\end{eqnarray}
The index $+$ denotes the usual distribution $F_+(z)=F(z)-\delta(1-z)\int_0^1
dz' F(z')$. \s

Including the vertex correction due to the squark-gluino exchange diagram and
the corresponding self-energy counterterm, the lowest order partonic 
cross section in eq.~(4) will be shifted by
\beq 
\hat{\sigma}_{\rm LO} \ra \hat{\sigma}_{\rm LO} \left[ 1+ \frac{2}{3} 
\frac{\alpha_s (\mu)}{\pi} \Re e C (\hat{s}, m_{\tilde{q}}, m_{\tilde{g}})
\right] 
\eeq
For degenerate unmixed squarks [as is approximately the case for the first two 
generation squarks], the expression of the factor $C$ is simply given by
\beq
C (s, m_{\tilde{q}}, m_{\tilde{g}})
= 2 \int_0^1 x{\rm d}x \int_0^1 {\rm d} y {\rm log} \frac{
m_{\tilde{g}}^2 +( m_{\tilde{q}}^2 -m_{\tilde{g}}^2)x}
{-s x^2 y(1-y) +( m_{\tilde{q}}^2 -m_{\tilde{g}}^2)x
+m_{\tilde{g}}^2 - i\epsilon}
\eeq
In terms of the Passarino--Veltman scalar functions $A_0, B_0$ and $C_0$
\cite{PV} the expression of the function $C$ reads
\beq
C (\hat{s}, m_{\tilde{q}}, m_{\tilde{g}})
&=& 2 m_{\tilde{g}}^2 C_0 (\hat{s}, 0, m_{\tilde{q}},m_{\tilde{q}},
m_{\tilde{g}}) - \frac{2}{\hat{s}} (m_{\tilde{q}}^2-m_{\tilde{g}}^2) C_+(\hat{s},0, 
m_{\tilde{q}}, m_{\tilde{q}},m_{\tilde{g}}) \non \\
&& +1 +B_0(\hat{s}, m_{\tilde{q}}, m_{\tilde{q}}) - 2B_1 (0, m_{\tilde{g}}, 
m_{\tilde{q}})
\label{eq:cfac}
\eeq
with 
\beq
C_+ (s, 0, m_1, m_1, m_2) &=& B_0(s, m_1, m_1) 
-B_0(0, m_1,m_2) + (m_2^2-m_1^2) C_0 (s, 0, m_1, m_1, m_2) \nonumber \\
B_1 (0, m_1, m_2)&=& \frac{1}{2} \left[ B_0 (0, m_1,m_2) + (m_1^2-m_2^2)
B'_0(0,m_1,m_2) \right]
\eeq
where
\beq
B'_0(s,m_1,m_2) = \frac{\partial}{\partial s} B_0(s,m_1,m_2)
\eeq
It should be noted that these results agree with the corresponding SUSY-QCD
corrections to slepton pair production at hadron colliders \cite{gaugino}.

\subsubsection*{2.2 The vector boson fusion processes}

The differential leading order partonic cross section for the vector boson 
fusion process, $q q \ra qqV^* V^* \ra q q \Phi$ with $V = W,Z$ and $\Phi=h,H$
[with the couplings $g_{\Phi VV}$ given before] can be cast into the form \cite{habil,vvh} 
\begin{eqnarray}
d\sigma_{LO} & = & \frac{1}{4} \frac{\sqrt{2}G_F^3M_V^8 q_1^2 q_2^2
g_{\Phi VV}^2}
{[q_1^2-M_V^2]^2 [q_2^2-M_V^2]^2} \left\{
F_1(x_1,M^2) F_1(x_2,M^2) \left[ 2+\frac{(q_1 q_2)^2}{q_1^2q_2^2} \right]
\right. \nonumber \\
&&+\frac{F_1(x_1,M^2)F_2(x_2,M^2)}{P_2q_2}\left[\frac{(P_2q_2)^2}{q_2^2}-
M_P^2+\frac{1}{q_1^2}\left(P_2q_1-\frac{P_2q_2}{q_2^2}q_1q_2\right)^2 \right]
\nonumber \\
&&+\frac{F_2(x_1,M^2)F_1(x_2,M^2)}{P_1q_1}\left[\frac{(P_1q_1)^2}{q_1^2}-
M_P^2+\frac{1}{q_2^2}\left(P_1q_2-\frac{P_1q_1}{q_1^2}q_1q_2\right)^2 \right]
\nonumber \\
& & +\frac{F_2(x_1,M^2)F_2(x_2,M^2)}{(P_1q_1)(P_2q_2)}\left[P_1P_2 -
\frac{(P_1q_1)(P_2q_1)}{q_1^2} - \frac{(P_2q_2)(P_1q_2)}{q_2^2} \right.
\nonumber \\
& & \left. \hspace*{4.5cm}+\frac{(P_1q_1)(P_2q_2)(q_1q_2)}{q_1^2q_2^2}\right]^2
\nonumber \\ 
& &\left. +\frac{F_3(x_1,M^2)F_3(x_2,M^2)}{2(P_1q_1)(P_2q_2)}\left[
(P_1P_2)(q_1q_2) - (P_1q_2)(P_2q_1) \right] \right\} dx_1 dx_2
\frac{dP\!S_3}{\hat s}
\label{eq:vvhlo}
\end{eqnarray}
where $dP\!S_3$ denotes the three-particle phase space of the final-state
particles, $M_P$ the proton mass, $P_{1,2}$ the proton momenta and $q_{1,2}$ the
momenta of the virtual vector bosons $V^*$.
The functions $F_i(x,M^2)~(i=1,2,3)$ are the usual structure functions from
deep-inelastic scattering processes at the factorization scale $M$:
\begin{eqnarray}
F_1(x,M^2) & = & \sum_q (v_q^2+a_q^2) [q(x,M^2) + \bar q(x,M^2)] \nonumber \\
F_2(x,M^2) & = & 2x \sum_q (v_q^2+a_q^2) [q(x,M^2) + \bar q(x,M^2)] \nonumber \\
F_3(x,M^2) & = & 4 \sum_q v_qa_q [-q(x,M^2) + \bar q(x,M^2)]
\label{eq:stfu}
\end{eqnarray}
where $v_q$ and $a_q$ have been defined previously. \s

In the past the standard QCD corrections have been calculated within the 
structure function approach \cite{vvhqcd}. Since at lowest order, the
proton remnants are color singlets, at NLO no color will be exchanged between
the first and the second incoming (outgoing) quark line and hence the QCD
corrections only consist of the well-known corrections to the structure
functions $F_i(x,M^2)~(i=1,2,3)$. The final result for the QCD-corrected
cross section can be obtained from the replacements

\begin{eqnarray}
F_i(x,M^2)  \to  F_i(x,M^2) + \Delta F_i(x,M^2,Q^2) \hspace*{1cm} (i=1,2,3)
\end{eqnarray}
with \cite{habil,vvhqcd}

\begin{eqnarray}
\Delta F_1(x,M^2,Q^2) & = & \frac{\alpha_s(\mu)}{\pi}\sum_q (v_q^2+a_q^2)
\int_x^1 \frac{dy}{y} \left\{ \frac{2}{3} [q(y,M^2) + \bar q(y,M^2)]
\right. \nonumber \\
& &
\left[ -\frac{3}{4} P_{qq}(z) \log \frac{M^2z}{Q^2} + (1+z^2) {\cal D}_1(z)
- \frac{3}{2} {\cal D}_0(z) \right. \nonumber \\
& & \left. \hspace*{6cm} + 3 - \left(
\frac{9}{2} + \frac{\pi^2}{3} \right) \delta(1-z) \right]
\nonumber \\
& & \left. + \frac{1}{4} g(y,M^2) \left[ -2 P_{qg}(z) \log \frac{M^2z}{Q^2(1-z)}
 + 4z(1-z) - 1 \right] \right\} \\
\Delta F_2(x,M^2,Q^2) & = & 2x\frac{\alpha_s(\mu)}{\pi}\sum_q (v_q^2+a_q^2)
\int_x^1 \frac{dy}{y} \left\{ \frac{2}{3} [q(y,M^2) + \bar q(y,M^2)]
\right. \nonumber \\
& &
\left[ -\frac{3}{4} P_{qq}(z) \log \frac{M^2z}{Q^2} + (1+z^2) {\cal D}_1(z)
- \frac{3}{2} {\cal D}_0(z) \right. \nonumber \\
& & \left. \hspace*{3.0cm} + 3 + 2z - \left(
\frac{9}{2} + \frac{\pi^2}{3} \right) \delta(1-z) \right]
\nonumber \\
& & \left. + \frac{1}{4} g(y,M^2) \left[ -2P_{qg}(z) \log \frac{M^2z}{Q^2(1-z)}
+ 8z(1-z) - 1 \right] \right\} \\
\Delta F_3(x,M^2,Q^2) & = & \frac{\alpha_s(\mu)}{\pi} \sum_q 4 v_q a_q
\int_x^1 \frac{dy}{y} \left\{ \frac{2}{3} [-q(y,M^2) + \bar q(y,M^2)]
\right. \nonumber \\
& &
\left[ -\frac{3}{4} P_{qq}(z) \log \frac{M^2z}{Q^2} + (1+z^2) {\cal D}_1(z)
- \frac{3}{2} {\cal D}_0(z) \right. \nonumber \\
& & \left. \left. \hspace*{3cm} + 2 + z - \left(
\frac{9}{2} + \frac{\pi^2}{3} \right) \delta(1-z) \right] \right\} \, ,
\end{eqnarray}
where $z=x/y$ and the Altarelli--Parisi splitting functions $P_{qq}, P_{qg}$ 
are as given before. We introduced the notation ${\cal D}_i(z) =
(\log^i(1-z)/(1-z))_+$ $(i=0,1)$. The physical scale $Q$ is given by
$Q^2 = -q_i^2$ for $x=x_i~(i=1,2)$.
These expressions have to be inserted in eq.~(\ref{eq:vvhlo}) and the full
result expanded up to NLO. The typical renormalization and factorization scales
are fixed by the corresponding vector-boson momentum transfer
$\mu^2=M^2=-q_i^2$ for $x=x_i$ ($i=1,2$). \s

The inclusion of the (two) vertex corrections due to the squark-gluino vertex 
diagrams and the corresponding self-energy counterterms, assuming that
the quarks are massless and the squarks degenerate in mass, can be performed 
by shifting the functions $F_i(x_j, M^2)$ in eq.~(13) by ($i=1,\ldots,3$ and $j=1,2$)
\beq
F_i(x_j, M^2) \ra F_i(x_j, M^2) \bigg[ 1+ \frac{2}{3} \frac{\alpha_s (\mu)}
{\pi} \Re e C (q_j^2, m_{\tilde{q}}, m_{\tilde{g}}) \bigg]
\eeq
where the function $C(\hat{s}, m_{\tilde{q}}, m_{\tilde{g}})$ has been 
given in eq.(\ref{eq:cfac}). 

\subsection*{3. Numerical Results} 

We will perform our numerical analysis for the light scalar Higgs boson $h$ in 
the decoupling limit of large pseudoscalar masses, $M_A \sim 1$ TeV. In this 
case the light $h$ boson couplings to standard particles approach the SM
values. In our analysis we will investigate the Higgs strahlung process 
$q\bar q \to hV$ and the vector boson fusion mechanism $qq\to qqV^*V^*\to qqh$ 
[$V=W,Z$]. \s

In Fig.~2, we show the cross sections for these processes at LO and 
at NLO with only the standard QCD corrections included, as a function of the 
$h$ boson mass [which can be made varied by varying the parameter $\tan \beta$]
for Tevatron and LHC energies. The NLO (LO) cross 
sections are convoluted with CTEQ4M (CTEQ4L) parton densities \cite{cteq4} and 
NLO (LO) strong couplings $\alpha_s$. As can be inferred from the figure, 
the standard QCD corrections increase the Higgs-strahlung cross sections by 
about 30\% (40\%) at the LHC (upgraded Tevatron) and the fusion 
process by about 10\% (5\%) at the LHC (upgraded Tevatron). \s

\begin{figure}[hbtp]
\vspace*{0.4cm}

\hspace*{-1.5cm}
\begin{turn}{-90}%
\epsfxsize=7cm \epsfbox{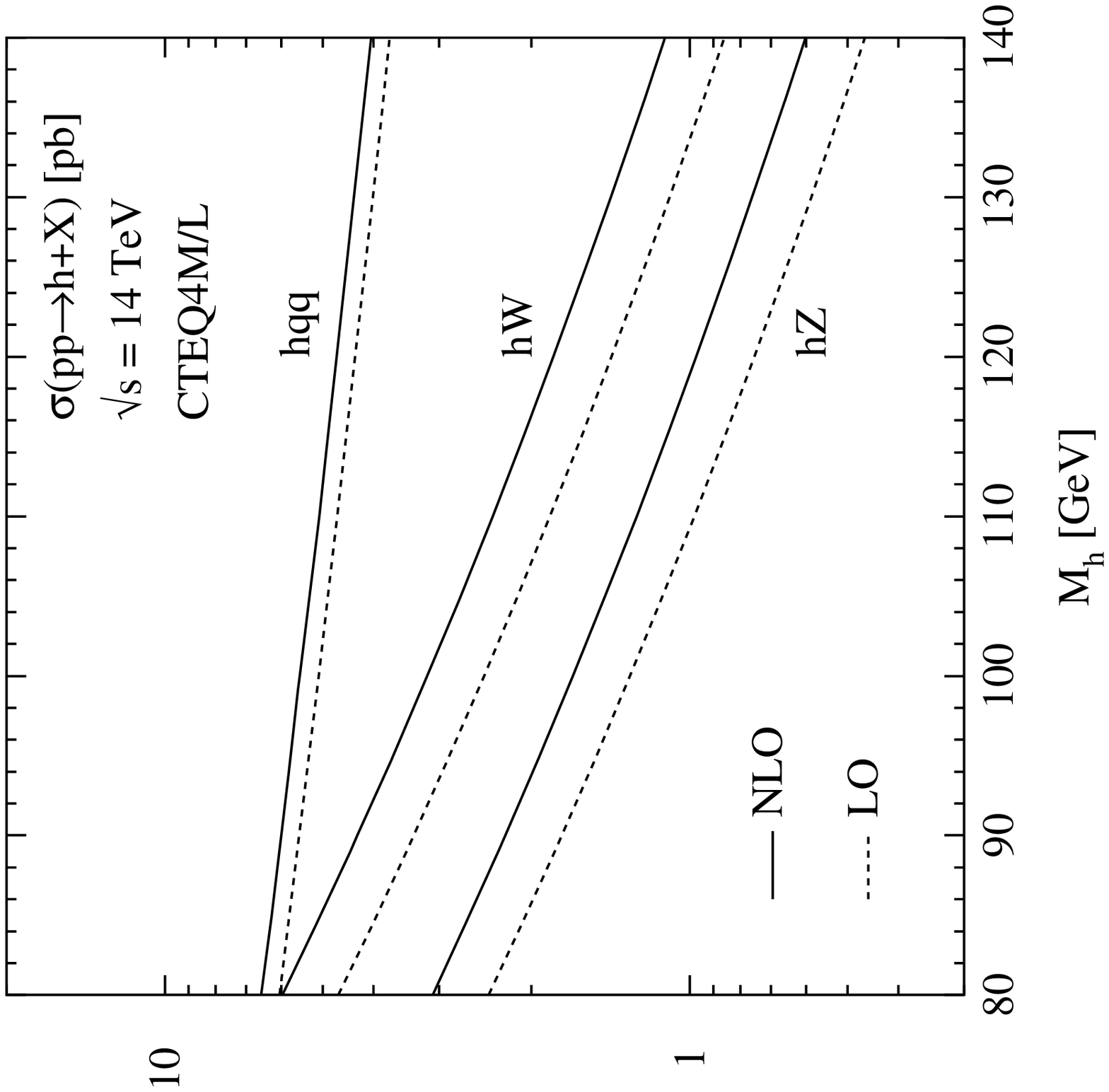}
\end{turn}
\hspace*{-1.8cm}
\begin{turn}{-90}%
\epsfxsize=7cm \epsfbox{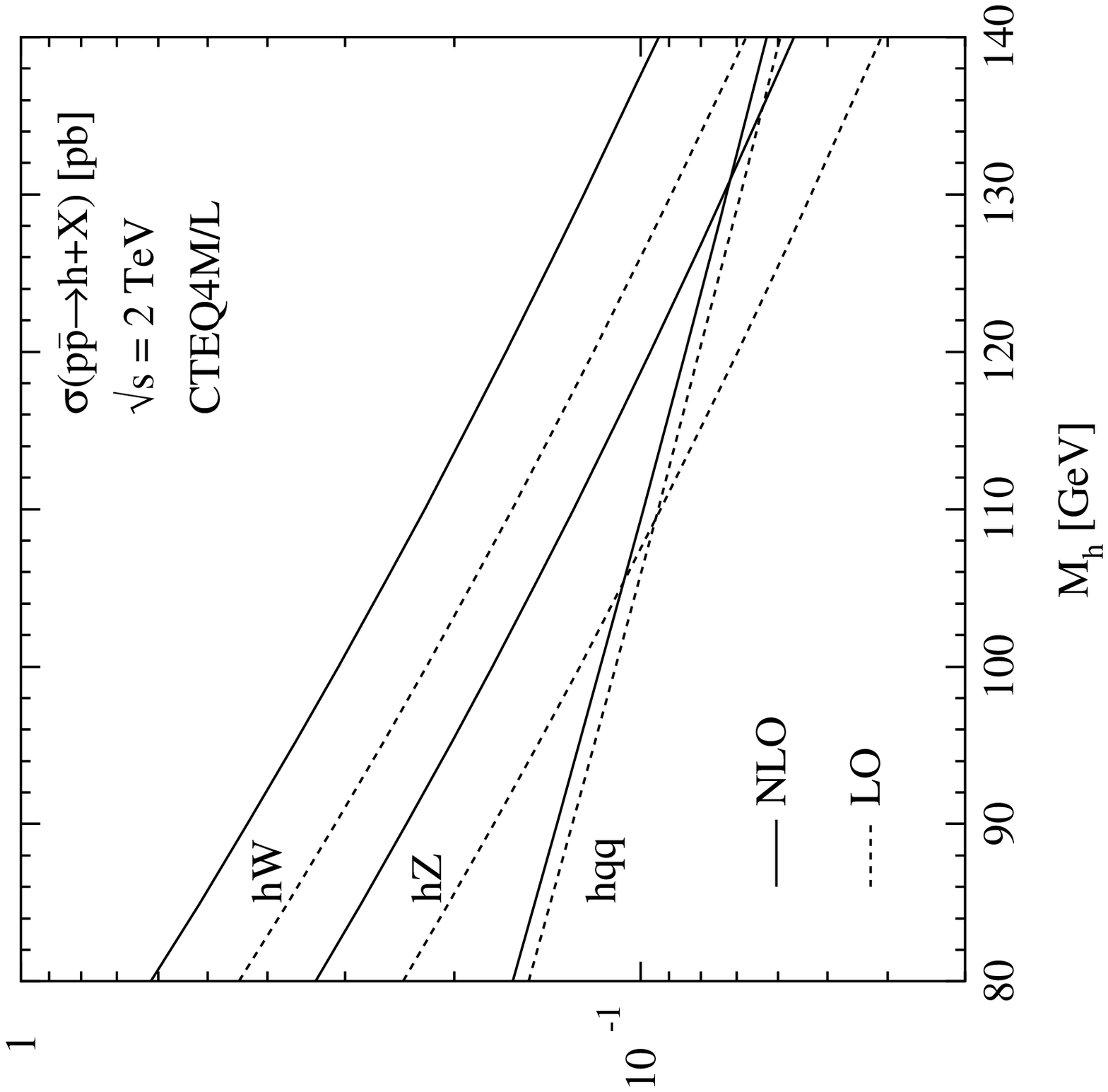}
\end{turn}
\caption[ ]{\it Total LO and NLO cross sections of Higgs boson production via
Higgs-strahlung $q\bar q\to h+W/Z$ and vector boson fusion $qq\to
qqV^*V^*\to qqh$ [$V=W,Z$] at the LHC (left) and the Tevatron (right) in
the decoupling limit.}
\label{fg:sigtot}
\end{figure}
\begin{figure}[hbtp]
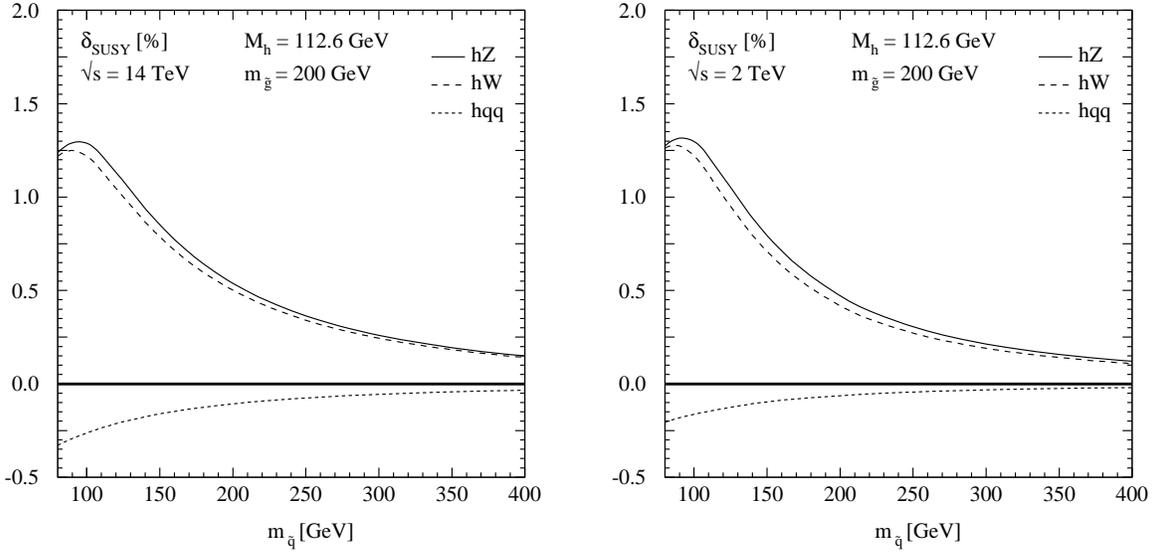

\vspace*{0.4cm}
\hspace*{-1.5cm}
\begin{turn}{-90}%
\epsfxsize=7cm \epsfbox{ksusy.lhc}
\end{turn}
\hspace*{-1.8cm}
\begin{turn}{-90}%
\epsfxsize=7cm \epsfbox{ksusy.tev}
\end{turn}
\caption[ ]{\it Relative corrections due to  virtual squark and
gluino exchange diagrams to Higgs boson production via Higgs-strahlung
$q\bar q\to h+W/Z$ and vector boson fusion $qq\to qqV^*V^*\to qqh$
[$V=W,Z$] at the LHC (left) and the Tevatron (right). }
\label{fg:susy}
\end{figure}

For the SUSY--QCD corrections, we evaluated the Higgs mass for $\tgb=30$, 
$M_A=1$ TeV and vanishing mixing in the stop sector; this yields a value 
$M_h=112.6$ GeV for the light scalar Higgs mass. For the sake of simplicity 
we decompose the $K$ factors $K=\sigma_{NLO}/\sigma_{LO}$ into the usual QCD 
part $K_{QCD}$ and the additional SUSY correction $\delta_{SUSY}$: $K = K_{QCD}
+\delta_{SUSY}$. The additional SUSY-QCD
corrections $\delta_{SUSY}$ are presented in Fig.~\ref{fg:susy} as a function 
of a common squark mass for a fixed gluino mass $m_{\tilde g}=200$ GeV [for
the sake of simplicity we kept the stop mass fixed for the determination of 
the Higgs mass $M_h$ and varied the loop-squark mass independently]. \s

The SUSY-QCD corrections increase the
Higgs-strahlung cross sections by less than 1.5\%, while they decrease the
vector boson fusion cross section by less than 0.5\%. The maximal shifts
are obtained for small values of the squark masses of about 100 GeV, which are 
already ruled out by present Tevatron analyses \cite{PDG}; for more reasonable 
values of these masses, the corrections are even smaller. Thus, the additional
SUSY-QCD corrections, which are of similar size at the LHC and the Tevatron,
turn out to be very tiny. For large squark/gluino masses they become even 
smaller due to the decoupling of these particles, as can be inferred from the 
upper squark mass range in Fig.~\ref{fg:susy}. 

\subsection*{4. Conclusions} 

In this letter we have determined the SUSY-QCD corrections to Higgs
boson production via Higgs-strahlung, Drell--Yan-like Higgs pair
production and vector boson fusion at hadron colliders. The additional
SUSY corrections originating from the exchange of virtual squarks and
gluinos range at the per cent level and are thus rather small. This
analysis completes the theoretical calculation of the NLO production
cross sections of these processes in the framework of supersymmetric
extensions of the Standard Model. 

\bigskip

\noindent
{\bf Acknowledgements.} \s

\noindent We thank P.M.~Zerwas for useful discussions.


\end{document}